\documentclass{nature}
\usepackage{epsfig}
\usepackage{amsmath}

\usepackage{amsmath,mathrsfs,amsbsy,color,graphicx,bm,amsthm,amsfonts}
\usepackage{units}
\usepackage{bbm}
\usepackage{times}
\usepackage{dcolumn}
\usepackage{mathrsfs}
\usepackage{amsmath,amssymb,epsfig,float}

\begin{document}

\title{Quantum metrology and estimation of Unruh effect}
\author{Jieci Wang$^{1,2,\star}$, Zehua Tian$^{1}$, Jiliang Jing$^{1}$, and Heng Fan$^{2,3,\star}$}
\maketitle

\begin{abstract}
We study the quantum metrology for a pair of entangled Unruh-Dewitt detectors when one of them is accelerated and coupled to a massless scalar field. Comparing with previous schemes, our model requires only local interaction and avoids the use of cavities in the probe state preparation process. We show that the probe state preparation and the interaction between the accelerated detector and the external field have significant effects on the value of quantum Fisher information, correspondingly pose variable ultimate limit of precision in the estimation of Unruh effect. We find that the precision of the estimation can be improved by a larger effective coupling strength and a longer interaction time. Alternatively, the energy gap of the detector has a range that can provide us a better precision. Thus we may adjust those parameters and attain a higher precision in the estimation.  We also find that an extremely high acceleration is not required in the quantum metrology process.
\end{abstract}

\begin{affiliations}
\item
Department of Physics, and Key Laboratory of Low
Dimensional, Quantum Structures and Quantum
Control of Ministry of Education,
 Hunan Normal University, Changsha, Hunan 410081, China.
\item
Beijing National Laboratory for Condensed Matter Physics, Institute
of Physics, Chinese Academy of Sciences, Beijing 100190, China.
\item
Collaborative Innovation Center of Quantum Matter, Beijing 100190, China

$^\star$e-mail:jieciwang@gmail.com; hfan@iphy.ac.cn

\end{affiliations}
It is well known that a uniformly accelerated detector which interacts with external fields becomes excited in the Minkowski vacuum.
This effect is named as Unruh effect \cite{unruh1976, unruhreview}, which indicates the fact that quantum properties of fields is observer dependent \cite{teleport,Schuller-Mann,Lamata,RQI1,RQI3,RQI4,
RQI5}. However, despite its crucial role in modern theoretical physics,  the experimental detection of the Unruh radiation remains an open research program on date. The main technical obstacle is that the Unruh temperature for the current experimental realizable acceleration lies far below the observable threshold of temperature. More specifically, the Unruh temperature is smaller than 1 Kelvin even for  accelerations
up to $10^{21} m/s^2$  \cite{unruhreview,RQIchen, RQI2}. On the other hand, quantum metrology \cite{advances} aims to study the bounds of the estimation precision and the quantum strategies that can attain them.
The estimation is based on measurements made on a probe system that undergoes an
evolution depending on the estimated parameters.
For a classical metrology scheme, the effect of statistical errors can
be reduced by repeating the measurements and averaging the outcomes.
Furthermore, by using some quantum resources and taking into account laws of quantum mechanics, the
precision can be enhanced. More specifically, the mean variance of the errors for a given measurement on the parameter $\theta$ is bounded by the Cram\'{e}r-Rao inequality \cite{Cramer:Methods1946} $Var(\theta) \geq  [n\mathcal{F}_{\xi}(\theta)]^{-1}$, where $n$ is the number of identical measurements repeated and
$\mathcal{F}_{\xi} (\theta)$ is the Fisher information (FI) for a given measurement scheme. Moreover,  by opti   mizing over all the possible set of quantum measurements, the ultimate limit on the variance is
set by the quantum Cram\'{e}r-Rao  bound $Var(\theta) \geq  [n\mathcal{F}_{Q}(\theta)]^{-1}$,
where $\mathcal{F}_{Q}(\theta)\geq \mathcal{F}_{\xi}(\theta)$ is the \emph{quantum} Fisher information (QFI). Recently, the adaptation of quantum metrology to improve probing technologies of relativistic effects has been preceded by
several pioneering works in different contexts,
see for example \cite{Aasi, Sabin,Doukas,aspachs,HoslerKok2013,RQI6,RQM,RQM2}.
These studies are of great
importance for the observation of relativistic effects in laboratories \cite{casimirwilson} and
pace-based quantum information processing tasks \cite{zeilingerteleport,bitcom,Bruschi1306}.

In particular, it is found that quantum metrology can be employed to enhance the accuracy of estimation for the Unruh effect, both for accelerated free modes \cite{aspachs,HoslerKok2013,RQI6} and moving cavities ~\cite{RQM,RQM2}.
Unfortunately, the former is suffered from the physically unfeasible  detection of global free models in the full spacetime while the latter is absence of non-perturbative expression of Boglivbov coefficients and therefore without an analytical form of QFI due to the boundary conditions of the moving cavities. To avoid these difficulties, in this paper we employ the Unruh-Dewitt detector model \cite{UW84} and avoid the use of any cavities. The detector is modeled by a two-level semiclassical atom with a fixed energy gap and interacts only with the neighbor field. We assume that one detector is switched off and at the same time keeps stationary while the other detector moves with constant acceleration and interacts with a massless scalar filed in the Minkowski vacuum. The detector is \emph{classical} in the sense that it possesses a classical world line while \emph{quantum} because its internal degree of freedom are treated quantum mechanically \cite{UW84}.
This model is adopted to study the behaviour of quantum teleportation \cite{Landulfo}, quantum discord \cite{Landulfo1} and quantum nonlocality \cite{tianwang} under the influence of the Unruh effect. We find that the strength of effective coupling between the accelerated detector and the external field, and the energy gaps of the detector have significant effects on the value of QFI, i.e., the precision in the
estimation of Unruh effect is sensitive to those parameters. We also find that an extremely high acceleration is not required in the estimation of the Unruh effect, although the higher Unruh temperature is obtained for a higher acceleration.

\section*{Results}
\subsection{Detector-Field interaction and probe state preparation}

We consider a couple of Unruh-Dewitt detectors \cite{UW84} in the Minkowski spacetime, each of them is modeled through a two-level non-interacting atom \cite{Landulfo,Landulfo1,tianwang}. The detectors initially share a entangled state, which is the probe state of the estimation and has the form as,
\begin{equation}
|\Psi_{AR}\rangle=\sin\theta |0_{A}\rangle|1_{R}\rangle+\cos\theta|1_{A}\rangle
|0_{R}\rangle,\label{IS}%
\end{equation}
where $|0_{A(R)}\rangle$ and $|1_{A(R)}\rangle$ represent the unexcited and excited states of Alice's (or Rob's) detector, respectively. The total initial state of the detectors plus the external scalar fields is given by $
|\Psi_{t_0}^{AR\phi}\rangle=|\Psi_{AR}\rangle\otimes|0_{M}\rangle$, where $|0_{M}\rangle$ corresponds to the Minkowski vacuum of the scalar field.

We assume that the detector carried by  Alice keeps inertial while  Rob's detector moves with constant acceleration $a$ along the $x$ axis for a finite amount of
time $\Delta$. The world line of a uniformly accelerated detector is described by
\[
t(\tau)=a^{-1}\sinh a\tau,\;x(\tau)=a^{-1}\cosh a\tau,
\]
$y(\tau)=z(\tau)=0$, where $\tau$ is the detector's proper time and $(t,x,y,z)$ are
Cartesian coordinates of the Minkowski spacetime. Throughout this paper we set $c=\hbar=\kappa_{B}=1$. From Alice's  perspective, the full Hilbert space is $\mathbf{H} = \mathbf{H}_A \otimes \mathbf{H}_R \otimes
\mathbf{H}_I$. However, we should map the state in the accelerated observer of Rob's frame into the Rindler Fock space basis, which means that a complementary Rindler region $\mathbf{H}_{II}$ is relevant. Here $\mathbf{H}_I$ is the Hilbert space in the right region of the Rindler spacetime, and analogously $\mathbf{H}_{II}$ denotes that in the left Rindler region.

Then we let
Alice's detector keeps switched off while Rob's detector is switched on at the very beginning of its accelerated motion. The detectors are two-level atoms with energy gap $\Omega$ as introduced by Unruh and Wald \cite{UW84}. Besides, the detectors are assumed to be point-like and only interact with the field in neighborhoods of their world lines. The total Hamiltonian of the  system is given by
\begin{equation}
H_{A\, R\, \phi} = H_A + H_R + H_{KG} +  H^{R\phi}_{\rm int},
\end{equation}
where $H_{A}=\Omega A^{\dagger}A$ and $H_{R}=\Omega R^{\dagger}R$ are the detectors' Hamiltonian, $H_{KG}$ is Hamiltonian of the massless scalar field, and $H^{R\phi}_{\rm int}$ is the interaction Hamiltonian between the scalar field and Rob's detector.
Rob's detector will keep being switched on for a finite amount of
time $\Delta$ and interacts with the external scalar field $\phi(x)$ which satisfies the massless Klein-Gordon equation
$\nabla_a \nabla^a \phi(x)=0$~\cite{indeces}.

The density matrix that describes the detector's state after the accelerated motion and the interaction is found to be
\begin{eqnarray}
\rho_{t}^{AR} &=& \alpha|\Psi_{AR}\rangle \langle \Psi_{AR}|
           + \beta|0_A\rangle|0_R\rangle \langle 0_A| \langle 0_R|
           \nonumber \\
           &+& \gamma|1_A\rangle|1_R\rangle \langle 1_A| \langle 1_R|,
\label{rhof2}
\end{eqnarray}
where $\Psi_{AR}$ is the initial state of the detectors and the parameters
$\alpha$, $\beta$ and $\gamma$ are found to be
\begin{eqnarray}
\alpha &=&
\frac{(1-e^{-\Omega/T})}{(1-e^{-\Omega/T})
+\sin^2 \theta \nu^2 + \nu^2\cos^2 \theta  e^{-\Omega/T}},
\nonumber \\
\beta &=&
\frac{\nu^2\sin^2 \theta} {(1-e^{-\Omega/T})
+ \nu^2\sin^2 \theta +\nu^2\cos^2 \theta e^{-\Omega/T}},
\nonumber \\
\gamma&=& \frac{\nu^2 \cos^2 \theta e^{-\Omega/T}}{(1-e^{-\Omega/T})
+ \nu^2 \sin^2 \theta +\nu^2\cos^2 \theta e^{-\Omega/T}},
\nonumber
\end{eqnarray}
respectively, and $T=a/2\pi$ is the Unruh temperature. For the sake of convenience, we have defined the effective coupling \cite{Landulfo, wald94}
\begin{equation}
\nu^{2}\equiv||\lambda||^{2}=\frac{\epsilon^{2}\Omega\Delta}{2\pi}%
e^{-\Omega^{2}\kappa^{2}}, \label{nu}%
\end{equation}
where $\epsilon$ is  the coupling constant, $\Delta$ is the time interval of the interaction, and $\Omega^{-1}\ll\Delta$ is required for the  validity of the above definition. In this paper the coupling constant is fixed as $\epsilon=2\pi\times10^{-3}$ \cite{Landulfo, Landulfo1} and the effective coupling is restricted to $\nu\ll1$ \cite{sabin1} for the validity of the perturbative approach.  From Eq. (3) we can see that the Unruh temperature $T$ is involved in the evolution of the probe state.
Now our main task is to optimize the estimation procedure
by maximizing the precision over all the interaction parameters. In the following, by employing the optimal probe preparation and tuning the interaction
parameters, we are going to seek the optimal strategy for the Unruh temperature estimation.

\subsection{Quantum Fisher information and metrology for the detector model}

We assume that the following process is repeated $n$ times: Alice's and Rob's detectors are prepared initially as $\Psi_{AR}$ in the inertial frame,
then we let Rob's detector be accelerated for a duration of
time $\Delta$ while Alice stays inertial. After the interaction period, a set of positive operator valued measurement (POVM)  is performed on the final state $\rho_{t}^{AR}$. For each interaction period, we can get an unbiased estimator $\xi$ for the Unruh temperature $T$.
According to the classical Cram\'{e}r-Rao inequality \cite{Cramer:Methods1946},
the mean variance of the error for this measurement scheme is $Var(T)\geq1/\mathcal{F}_{\xi}(T)$, where $\mathcal{F}_{\xi}(T)$ is the FI for the estimated parameter T. The $\mathcal{F}_{\xi}(T)$ is defined as
\begin{align}
\mathcal{F}_{\xi}(T) = \Sigma_\xi p(\xi\vert \lambda) \left(
\frac{\partial \ln p(\xi\vert T)}{\partial T}
\right)^2
=\Sigma_\xi \frac{1}{p(\xi\vert T)} \left(
\frac{\partial p(\xi\vert T)}{\partial T}
\right)^2, \label{eq:ClassicalFisher}
\end{align}
where $p(\xi\vert T)$ is the probability of
obtaining the value $\xi$ when the parameter $T$ is estimated.
In quantum mechanics, according to the Born rule we have
$p(x\vert T) = Tr[\Pi_{\xi} \rho_{t}^{AR}]$,
where
$\rho_{t}^{AR}$ is the density operator of the final state. Now we define the symmetric logarithmic derivative (SLD) $\mathcal{L}_T$ as
\begin{align}
\frac{\mathcal{L}_T \rho_{t}^{AR} + \rho_{t}^{AR} \mathcal{L}_T}{2} =
\frac{\partial \rho_{t}^{AR}}{\partial T},
\end{align}\label{eq:SLD}
where the relation
$ \frac{\partial p(\xi\vert T)}{\partial T} = Tr[\frac{\partial (\rho_{t}^{AR} \Pi_\xi)}{\partial T}]
= \hbox{Re}( Tr[\rho_{t}^{AR} \Pi_\xi L_T ] )$ is used .
The FI (\ref{eq:ClassicalFisher}) is then can be rewritten as
\begin{align}
\mathcal{F}_{\xi}(T) = \Sigma_\xi \bigg[ \frac{\hbox{Re}\left(Tr\left[\rho_{t}^{AR} \Pi_\xi
\mathcal{L}_T \right]\right)^2} {Tr[\rho_{t}^{AR} \Pi_\xi]}\bigg]
\:. \label{eq:CQFisher}
\end{align}
For any given POVM $\{\Pi_\xi\}$, FI establish the
bound on precision. To obtain the ultimate bounds on
precision, now the task is maximizing the FI over all
the possible quantum measurements.
Following Refs. \cite{Braunstein1994, Braunstein1996}, we have
\begin{align}
\mathcal{F}_{\xi}(T)  & \leq
\Sigma_\xi \left|\frac{Tr\left[\rho_{t}^{AR} \Pi_\xi
L_T\right]}{\sqrt{Tr[\rho_{t}^{AR} \Pi_\xi]}}\right|^2\nonumber
\\
& \leq \ \Sigma_\xi Tr\left[\Pi_\xi \mathcal{L}_T \rho_{t}^{AR} \mathcal{L}_T\right]\nonumber
\\
& =Tr[\mathcal{L}_T \rho_{t}^{AR} \mathcal{L}_T ],
\end{align}
where the last term is the  QFI
\begin{align}
\mathcal{F}_{Q}(T)
= Tr[\partial_T \rho_{t}^{AR} \mathcal{L}_T]= Tr[\rho_{t}^{AR} \mathcal{L}_T^2].
\label{eq:QuantumFisher}
\end{align}
Thus,  optimizing over all the possible measurements provides us with an lower quantum Cram\'{e}r-Rao bound~\cite{Braunstein1994}, i.e.,
\begin{equation}\label{Cramer-Rao}
 Var(T)\geq\frac{1}{n\mathcal{F}_{\xi}(T)}\geq \frac{1}{n\mathcal{F}_{Q}(T)}.
\end{equation}
Despite the concise definition of
QFI, the calculation of $\mathcal{L}_T$ is somewhat a non-trival
task.  Alternatively, basing on a spectrum decomposition of the state as $\rho_{t}^{AR}=\sum_{m=1}^N p_m|\psi_m\rangle\langle\psi_m|$,
the QFI can be rephrased as  \cite{Braunstein1996,Pairs2009}
\begin{equation}
\mathcal{F}_{Q}(T)=2\sum_{m,n}^N\frac{|\langle\psi_m|
\partial_T\rho_{t}^{AR}|\psi_n\rangle|^2}{p_m+p_n},
\end{equation}
with the eigenvalues $p_m\geq0$ and $\sum_m^Np_m=1$.
For a non-full-rank state the QFI can be
expressed as\cite{Zhong,Zhang2013}
\begin{eqnarray}
\nonumber\mathcal{F}_{Q}(T)&=&
  \sum_{m'}\frac{\left(\partial_{T}p_{m'}\right)^{2}}{p_{m'}}+\\
  &&2\sum_{m\neq n}\frac{\left(p_{m}-p_{n}\right)^{2}}{p_{m}+p_{n}}
  \left|\left\langle\psi_{m}|\partial_{T}\psi_{n}\right\rangle \right|^{2},
  \label{mixed}
\end{eqnarray}
where the summations involve sums over all $p_{m'}\neq0$ and
$p_{m}+p_{n}\neq0$, respectively.
From Eq.~\eqref{mixed}, we can see that the QFI of a non-full-rank state can be determined by the subset of the spectrum decomposition of the state with nonzero eigenvalues.

Our aim is to study how precisely one can in principle estimate
the Unruh temperature that appears in the detector model. We are looking for the optimal estimation scheme, i.e. finding the
optimal probe state preparation and interaction parameters that allow us to get the largest QFI. With the expression of the QFI in Eq. (\ref{mixed}), we only need the nonzero eigenvalues of the final state Eq. (\ref{rhof2}), there are
\begin{eqnarray}
\Lambda_1 &=&
\frac{2(1-e^{-\Omega/T})}{2-\nu^2(1+\cos 2\theta)+e^{-\Omega/T}[\nu^2(\cos2\theta-1)-2]
},
\label{s0}
\nonumber \\
\Lambda_2 &=&
\frac{\nu^2 \cos^2 \theta}{-1
+ \nu^2 \cos^2 \theta +e^{\Omega/T}(1
+ \nu^2 \sin^2 \theta)},
\label{s1}
\nonumber \\
\Lambda_3&=& \frac{e^{\Omega/T}\nu^2 \sin^2 \theta}{-1
+ \nu^2 \cos^2 \theta +e^{\Omega/T}(1
+ \nu^2 \sin^2 \theta)}.
\label{s2}
\nonumber
\end{eqnarray}
The corresponding eigenvectors are found to be $|\psi_1\rangle=(1+\tan^2 \theta)^{-1} \left\{0,\tan\theta,0,1\right\}$,
$|\psi_2\rangle=\left\{0,0,0,1\right\}$, and
$|\psi_3\rangle=\left\{1,0,0,0\right\}$, respectively. Now we have obtained all the required elements to calculate the QFI for the estimation of the Unruh temperature $T$. Physically, a fix value of FI is obtained by any set of measurement, while the QFI is the biggest FI optimizing over all the possible measurements. Here the eigenvectors $|\psi_m\rangle$ of the final state Eq.(3) independent of the estimated parameter $T$ so $\partial_T |\psi_m\rangle=0$. Then the optimal projective measurement can be constituted by the eigenvectors $|\psi_m\rangle$ of the final state, and the measured probabilities $p(\xi\mid T)$ are exactly the eigenvalues $p_m$ of the final state.

In Figure 1 we plot the QFI of the probe state Eq. (\ref{rhof2}) after the Unruh temperature involved evolution as functions of the effective coupling parameter $\nu$ and  the acceleration $a$.  The maximal QFI is obtained by numerical optimization over $\nu$ and $a$ for a given initial state parameter $\theta=\pi/4$, i.e., the initial between Alice's and Rob's detector is a singlet state. It is shown that the QFI always increase as the growth of coupling parameter $\nu$, which means that we can get the a larger precision for a stronger effective coupling between Rob's detector and the scalar field. We can see that the QFI of the final states depends on the observers's acceleration sensitively, which shows that the magnitude of Rob's acceleration has a non-trivial  influence on the quantum metrology. Note that the QFI firstly increases and then decreases as the increase of $a$, which indicates that the highest precision in the Unruh effect estimating can be obtained for a medium value acceleration. That is to say, we don't need to obtain an extremely high acceleration during the estimation of the Unruh effect, although the higher Unruh temperature is obtained for a higher acceleration. There is a range of acceleration that provides us with the optimal precision during the estimation procedure.

To obtain a physical interpretation of this counterintuitive phenomenon, we calculate the quantum entanglement of the final state Eq. (\ref{rhof2}).
We employ the well accepted concurrence \cite{Wootters,Coffman} to quantify quantum entanglement, which can be computed by $C(\rho)=2\max\left\{  0,\tilde{C}_{1}(\rho),\tilde{C}_{2}(\rho)\right\}$
 for a state with a $X$-type structure.
Here $\tilde{C}_{1}(\rho)=\sqrt{\rho_{14}\rho_{41}}-\sqrt{\rho_{22}\rho
_{33}}$ and $\tilde{C}_{2}(\rho)=\sqrt{\rho_{23}\rho_{32}}-\sqrt{\rho
_{11}\rho_{44}}$, and $\rho_{ij}$ are elements of the density matrix $\rho_{t}^{AR}$ of the final state.

In Figure 2 we compare the QFI and the entanglement of the probe state as a function of the acceleration $a$ for a fixed effective coupling parameter $\nu=0.1$. It is found that, as we have shown in Ref. \cite{tianwang}, the entanglement of the probe state decreases as the increasing of acceleration.
On the other hand, in order to have a detectable Unruh effect,
the acceleration is required to be large enough. This forms a balance:
it will be beneficial for the detecting of Unruh effect by a increasing acceleration,
but this large acceleration will destroy the entanglement
which in general will induce the increasing QFI \cite{advances} and thus
obscure overall the detection.
In Figure 2, we may notice that
the estimation precision increases rapidly as the increase of acceleration for small value accelerations
until reaching the optimal point, then decreases even the acceleration increases.
For some larger acceleration, the QFI diminishes due to the decrease of quantum entanglement. This suggests the regime of acceleration $a$ which is beneficial for the detection of Unruh effect.

We are also interested in how the energy gap $\Omega$ of Rob's detector and the interaction time $\Delta$ influence the estimation of the Unruh effect. In Figure 3 we plot the behaviour of QFI as a function of the energy gap $\Omega$ for different interaction time $\Delta$. It is shown that the QFI is sensitive to the variation of different energy gaps of Rob's detector. In particular, the maximal QFI is obtained at a fixed energy gap $\Omega$ value for every interaction time $\Delta$. That is to say, the energy gap of Rob's detector has a significant impact on the estimation of Unruh effect. Thus one can prepare a proper detector by some kinds of two-level systems that possess the proper  energy gap to obtain the best estimation precision. Alternatively, we can get a higher precision, i.e., a larger QFI for a longer interaction time. To sum up, we can choose the largest effective coupling strength and the longest interaction time allowed
by quantum mechanics, as well as some suitable energy gaps to realizes the optimal strategy attaining the ultimate sensitivity for the estimation of the Unruh effect.

\section*{Discussion}
We have studied the relativistic quantum
metrology for two entangled detectors when one of them with
accelerated motion. The optimal strategy for the Unruh effect estimation is obtained by employing the proper probe state preparation and by adjusting the interaction parameters in the estimation process. We employ the Unruh-Dewitt detector model, which has a fixed energy gap and interacts only with the neighbor field. The studied model avoids two critical technical difficulties in the estimation of the Unruh temperature: a physically unfeasible detection of global free models in the full space and a non-analytical expression of QFI due to the boundary conditions of the moving cavity. In this paper the point-like detectors only couple to the neighbour field modes, and Alice's and Rob's detectors in the Rindler region $I$ are causally connected. The studied modes are in fact relativistic local and only the local projective measurements are performed in the metrology process so obeys the causality. It is worthy to mention that the relativistic causality would be violated if the projective measurements are performed between the causally separated modes \cite{Lin0}. Fortunately, Lin recently found that the violation can be suppressed by introducing restrictions on the post-measurements for the projective measurements on relativistic nonlocal modes \cite{Lin1}. It is shown that the probe state preparation and the interaction parameters between Rob's detector and the external field have significant influences on the value of QFI. To be specific, there are a range of energy gaps of the accelerated detector that provide us a better precision in the estimation of the Unruh temperature. However, one should choose the largest effective coupling  strength and the physically allowed longest interaction time to achieve the same goal. The results of this paper can be also applied to the estimation of Hawking temperature of black holes and Unruh temperature for non-uniformly accelerated detectors \cite{hujhep}. Such topics are left for a future research.

\begin{methods}

The interaction Hamiltonian between Rob's detector and the scalar field is
\begin{equation}
H^{R\phi}_{\rm int}(t)=
\epsilon(t) \int_{\Sigma_t} d^3 {\bf x} \sqrt{-g} \phi(x) [\psi({\bf x})R +
                           \overline{\psi}({\bf x})R^{\dagger}],
\label{int}
\end{equation}
where $g\equiv {\rm det} (g_{ab})$, and $g_{ab}$ is the metric tensor of the Minkovski spacetime. Here $\epsilon$ is the coupling constant.  The detector is switched on smoothly within a finite time interval $\Delta$ and then switched off outside this interval. Besides, $\psi({\bf x})$ is a function which vanishes outside a small volume
around the detector, models the fact that the detector only interacts with the neighbor fields \cite{KY03} in the Minkowski vacuum.

The state $|\Psi^{R \phi}_{t = t_0+\Delta} \rangle$ that describes Rob's detector and the scalar field at time $t=t_0+\Delta$ can be expressed as
\begin{equation}
|\Psi^{R \phi}_t \rangle =
\mathbf{T} \exp[-i\int_{-\infty}^t dt' H_{\rm int} ^I(t')] |\Psi^{R \phi}_{t_0} \rangle,
\label{Dyson1}
\end{equation}
in the interaction picture, where $\mathbf{T}$ is the time-ordering operator and
\begin{equation}
H_{\rm int}^I (t) = U^{\dagger}_0(t) H_{\rm int} (t) U_0 (t).
\end{equation}
Here $U_0 (t)$ is an unitary evolution operator associated with
$H_R+H_{KG}$ \cite{Landulfo,wald94}. By using Eq.~(\ref{Dyson1}), we write the final state $|\Psi^{R \phi}_{t} \rangle$ of the detector-field system as
\begin{equation}
|\Psi^{R \phi}_{t} \rangle =
\mathbf{T} \exp [-i\int d^4x\sqrt{-g}\phi(x) (fR + \overline{f}R^{\dagger})] |\Psi^{R \phi}_{t_0} \rangle,
\label{Dyson2}
\end{equation}
where
$f \equiv \epsilon(t) e^{-i\Omega t}\psi ({\bf x})$
is a compact support complex function defined in the Minkowski spacetime.
In this paper we only consider the point-like detectors, which can be realized by choosing
$\psi(\mathbf{x})=(\kappa\sqrt{2\pi})^{-3}\exp(-\mathbf{x}^{2}/2\kappa^{2})$
with the parameter $\kappa=\mathrm{const}\ll 1$.
In the weak coupling case, we can express Eq.~(\ref{Dyson2}) in the first order of perturbation over the coupling constant $\epsilon$ \cite{UW84,Landulfo,Landulfo1}
\begin{equation}
|\Psi^{R \phi}_{t} \rangle
= [I - i(\phi(f)R + \phi(f)^{\dagger} R^{\dagger}) ] |\Psi^{R \phi}_{t_0} \rangle,
\label{primeira_ordem}
\end{equation}
where $\phi(f)$ is an operator valued distribution of the scalar field \cite{wald94} given by
\begin{eqnarray}
\nonumber\phi(f) &\equiv& \int d^4 x \sqrt{-g}\phi(x)f\\&=&i [a_{RI}(\overline{KE\overline{f}})-a_{RI}^{\dagger}(KEf)],
\label{phi(f)}
\end{eqnarray}
 and
$a_{RI}(\overline{u})$ and $a_{RI}^{\dagger}(u)$ are the annihilation and creation
operators of $u$ modes \cite{Landulfo,wald94}, respectively.
Besides, $K$ is an operator that takes the positive frequency part of the solutions of the
Klein-Gordon equation, and $E$ is the
difference between the advanced and retarded Green functions.

Considering that only Rob's detector interacts with the field, we evolve our initial state to its asymptotic form
\begin{eqnarray}
| \Psi^{AR \phi}_{t}\rangle
& = &
|\Psi^{AR \phi}_{t_0} \rangle
 + \sin \theta |0_A\rangle  |0_R\rangle
 \otimes(a_{R I}^{\dagger}(\lambda)|0_M\rangle)
 \nonumber \\
& + & \cos \theta |1_A\rangle |1_R\rangle\otimes(a_{R I}(\overline{\lambda})|0_M\rangle),
\label{evolutionAUX}
\end{eqnarray}
where  $\lambda = -KEf$, the subscripts in $a_{R I}^{\dagger}$ and $a_{R I}$ indicate that they are creation and annihilation operators of Rindler
modes in the region $I$. Note that in Eq. (\ref{evolutionAUX}) the
operators are defined in the Rindler coordinate, while the state $|0_M\rangle$ is
a vacuum state in the Minkowski spacetime.

We write the operators $a_{R I}$ and $a^{\dagger}_{R I}$
as~
\begin{eqnarray}
a_{R I}(\overline{\lambda})&=&
\frac{a_M(\overline{F_{1 \Omega}})+
e^{-\pi \Omega/a} a_M ^{\dagger} (F_{2 \Omega})}{(1- e^{-2\pi\Omega/a})^{{1}/{2}}},
\label{aniq} \\
a^{\dagger}_{R I}(\lambda)&=&
\frac{a^{\dagger}_M (F_{1 \Omega}) +
e^{-\pi \Omega/a}a_M(\overline{F_{2 \Omega}})}{(1- e^{-2\pi\Omega/a})^{{1}/{2}}}
\label{cria},
\end{eqnarray}
where
$F_{1 \Omega}=
\frac{\lambda+ e^{-\pi\Omega/a}\lambda\circ w}{(1- e^{-2\pi\Omega/a})^{{1}/{2}}}$, and
$F_{2 \Omega}=
\frac{\overline{\lambda\circ w}+ e^{-\pi\Omega/a}\overline{\lambda}}{(1- e^{-2\pi\Omega/a})^{{1}/{2}}}$. Here
$w(t, x)=(-t, -x)$
is the wedge reflection isometry, which makes a reflection from $\varphi\in \mathbf{H}_I$ to $\varphi\circ w \in \overline{\mathbf{H}}_{II}$.
It is worthy to note that the transformations Eqs. (20) and (21) are not the usual manner of the Bogoliubov transformations under the single-mode approximation. They are in fact the appropriate transformations \cite{UW84} between a set of positive-frequency modes $\lambda$ and $\lambda\circ w$ which are wave packet with frequencies peaked sharply about $\Omega$ and a set of functions $F_{1\Omega}$ and $F_{2\Omega}$ therefore beyond the single mode approximation.   Substituting the operators $a_{R I}(\lambda)$ and $a^{\dagger}_{R I}(\lambda)$ in Eq. (\ref{evolutionAUX}), the final state of the total system can be obtained.
The density matrix that describes the detector's state is calculated by tracing out the degrees of freedom of the external field
\begin{equation}
\rho^{AR}_{t}=\parallel  \Psi^{AR \phi}_{t}\parallel^{-2} Tr_{\phi} \Psi^{AR \phi}_{t}\rangle\langle\Psi^{AR \phi}_{t}|,\label{ddddd}
\end{equation}
where $\parallel | \Psi^{AR \phi}_{t}\parallel^{2}$ normalizes the final state and has the form of
\begin{equation}
\parallel  \Psi^{AR \phi}_{t}\parallel^{2}=1+\frac{\sin^2\theta \nu^2}{1- e^{-2\pi\Omega/a}}+\frac{\cos^2\theta \nu^2 e^{-2\pi\Omega/a}}{1- e^{-2\pi\Omega/a}}.
\end{equation}
Eq. (3) can be derived by working out Eq. (\ref{ddddd}).

\end{methods}

\begin{addendum}

\item [Acknowledgement]

This work is supported by 973 program through
2010CB922904, the National Natural Science Foundation
of China under Grant No. 11305058, No. 11175248, No. 11475061, the Doctoral Scientific Fund Project of the Ministry of Education of China under Grant No. 20134306120003, and Postdoctoral Science
Foundation of China under Grant No. 2014M560129.

\item [Author Contributions]
J. W. made the main calculations.
J. W., Z. T., J. J., and H. F. discussed the results, J. W. wrote the paper with assistances from H. F. and other authors.

\item [Competing Interests]
The authors declare that they have no competing financial interests.

\item [Correspondence]
Correspondence and requests for materials should be addressed to
J. W or H. F.
\end{addendum}

\newpage

\textbf{Figure 1. QFI in the estimation of the Unruh temperature as functions of the coupling parameter $\nu$ and the acceleration $a$.}
The initial state parameter is fixed with $\theta=\pi/4$ and the energy gap is given by $\Omega=1$. Here we set a smaller energy gap than that of Ref.\cite{Landulfo} because a smaller energy gap $\Omega$ makes the detector easier to be excited and de-excited by considering the metrology process repeat the measurement many times.

\textbf{Figure 2. QFI and entanglement of the final state Eq. (\ref{rhof2}) as a function of the acceleration $a$.} The initial state parameter is fixed with $\theta=\pi/4$ and the energy gap is given by $\Omega^{-1}=2\pi$. The effective coupling parameter is fixed as $\nu=0.1$ to keep the perturbative approach valid for large times.

\textbf{Figure 3. QFI in the estimation of the Unruh temperature $T$ as a function of the energy gap $\Omega$ for different interaction time $\Delta$.}
The parameters related to the effective coupling parameter are fixed to satisfy $\epsilon\ll\Omega^{-1}\ll\Delta$. They are fixed with $\epsilon=2\pi\cdot 10^{-3}$ and $\kappa=0.02$, respectively. The initial state parameter is given by $\theta=\pi/4$ and the acceleration parameter is fixed with $a=0.4\pi$.

\newpage
\begin{figure}[tbp]
\begin{center}
\epsfig{file=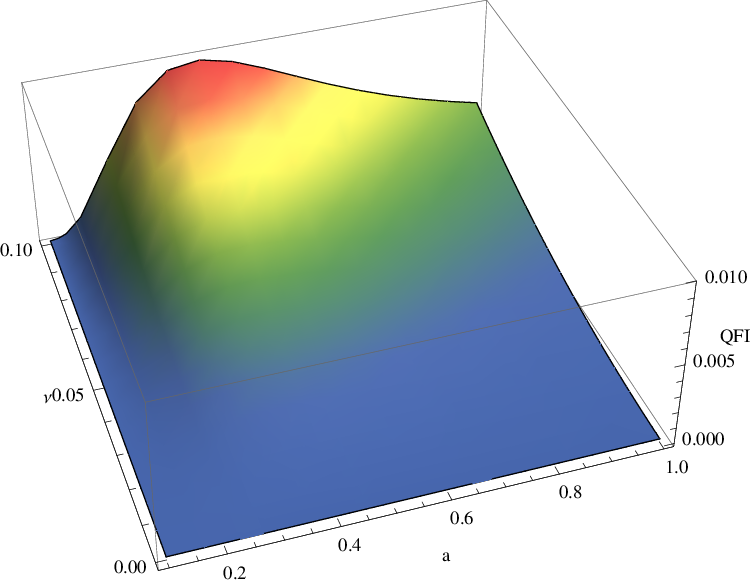,width=16cm}
\end{center}
\par
\label{Figure1}
\end{figure}

\begin{center}
Figure 1
\end{center}

\newpage

\begin{figure}[tbp]
\begin{center}
\epsfig{file=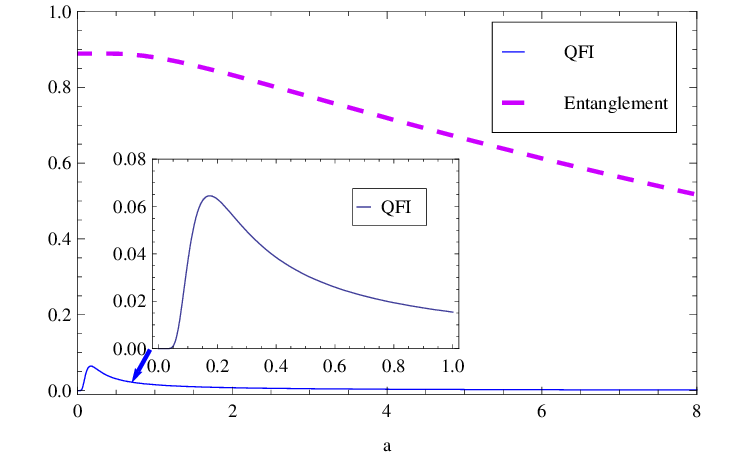,width=16cm}
\end{center}
\par
\label{Figure2}
\end{figure}

\begin{center}
Figure 2
\end{center}

\newpage


\begin{figure}[tbp]
\begin{center}
\epsfig{file=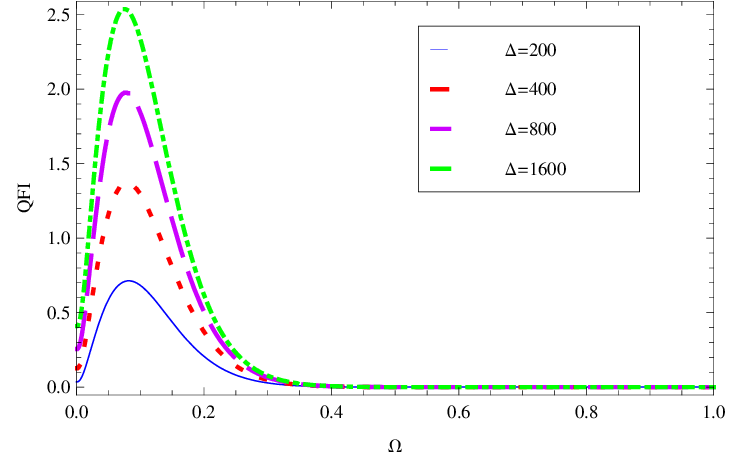,width=16cm}
\label{Figure3}
\end{center}
\end{figure}

\begin{center}
Figure 3
\end{center}

\newpage

\end{document}